\documentclass[sigconf]{acmart}

\renewcommand\footnotetextcopyrightpermission[1]{} 
\makeatletter
\renewcommand\@formatdoi[1]{\ignorespaces}
\makeatother
\acmConference[]{}{}{}

\AtBeginDocument{%
  \providecommand\BibTeX{{%
    \normalfont B\kern-0.5em{\scshape i\kern-0.25em b}\kern-0.8em\TeX}}}

\usepackage{multirow}
\usepackage{booktabs}
\usepackage{tabularx}

\begin{document}

\title{Reputation Agent: Prompting Fair Reviews in Gig Markets}
\author{Carlos Toxtli, Angela Richmond-Fuller}
\affiliation{%
  \institution{HCI Lab, West Virginia University}}
\email{carlos.toxtli@mail.wvu.edu,angela.richmond@gmail.com}

\author{Saiph Savage}
\affiliation{%
 \institution{Universidad Nacional Autonoma de Mexico (UNAM)}
 }
\email{saiph.savage@mail.wvu.edu}

\renewcommand{\shortauthors}{Carlos Toxtli, et al.}

\begin{abstract}
Our study presents a new tool, Reputation Agent, to promote fairer reviews from requesters (employers or customers) on gig markets. Unfair reviews, created when requesters consider factors outside of a worker's control, are known to plague gig workers and can result in lost job opportunities and even termination from the marketplace. Our tool leverages machine learning to implement an intelligent interface that: (1) uses deep learning to automatically detect when an individual has included unfair factors into her review (factors outside the worker's control per the policies of the market); and (2) prompts the individual to reconsider her review if she has incorporated unfair factors. To study the effectiveness of Reputation Agent, we conducted a controlled experiment over different gig markets. Our experiment illustrates that across markets, Reputation Agent, in contrast with traditional approaches, motivates requesters to review gig workers' performance more fairly. We discuss how tools that bring more transparency to employers about the policies of a gig market can help build empathy thus resulting in reasoned discussions around potential injustices towards workers generated by these interfaces. Our vision is that with tools that promote truth and transparency we can bring fairer treatment to gig workers.

\end{abstract}

\maketitle

\section{Introduction}

\begin{figure}
\centering
  \includegraphics[width=0.9\columnwidth]{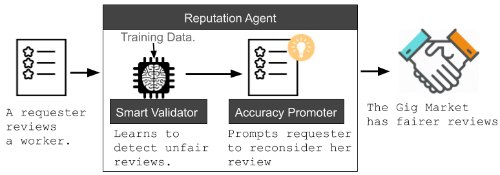}
  \caption{Overview of how Reputation Agent functions.}
  \label{fig:img01}
\end{figure}

Gig marketplaces are online spaces where almost anyone can contract independent workers (e.g., freelancers) to conduct labor or deliver services in the form of short-term engagements \cite{de2015rise}. Gig markets facilitate transactions between strangers, as people typically have to hire and manage a crowd of workers they have never met \cite{todoli2017end,bederson2011web}. Similarly, workers on these platforms often coordinate with other workers\cite{irani2013turkopticon} and offer services to requesters (customers or employers) who are all unfamiliar to them \cite{prassl2018humans}. These direct interactions between strangers mean that gig markets must have mechanisms through which people can truthfully assess each other, i.e., earning money by entrust their hard labor to strangers \cite{uslaner1999trust, qiu2018more, silberman2017rating,felstiner2011working,mcinnis2016taking}. One of the most popular instruments for helping people to assess strangers and to choose who to hire are reputation systems which function by asking individuals to provide feedback on others' work. These systems are generally based on the platform's review metrics.
For gig worker’s, reputation has become especially important because it is critical for accessing higher-paying jobs \cite{martin2014being, ma2018using} or even staying employed \cite{gray2019ghost}.

In this context, it is important to understand that in the power dynamics of most gig markets, the platform takes the side of requesters \cite{todoli2017end} or the platforms manipulate the market to the detriment of the worker \cite{calo2017taking,todoli2017end}. Therefore, if a requester invests time to write a lengthy complaint about a worker (even if the requester is incorrect), the market will side with the requester, potentially leading to unfair termination. For instance, the following advise is from an Uber driver for other gig drivers \cite{FiredFro8:online}: \emph{``...[one of the main reasons for Uber to terminate a driver is that the] passenger makes a serious complaint about you [the driver]. If a passenger goes out of their way to tell Uber that you were rude, or that you're a bad driver, or that you made them uncomfortable in any way, you can be immediately deactivated without prior notice. You aren't likely to be reactivated after a major passenger complaint…''} 

This environment where workers have limited mechanisms to negotiate or even discuss reviews has led workers to distrust gig markets altogether \cite{gray2019ghost, horton2015reputation}.
Thus, it is crucial to ensure that the reviews about workers are fair in order to improve trust and the general operation of gig markets \cite{beal2009radically}. 
Fairness within gig markets typically involves ensuring that the policies on the market are transparent, concise, and accessible to workers \cite{graham2018towards,silberman2009fifteen}. However, we argue that fairness is not just about empowering workers to understand the policies of the markets in which they participate; it is also about guiding the requester in the market to evaluate workers based on the market's established policies \cite{fan2005evaluation}.

Requesters should have a clear understanding of what metrics they should consider and which are inappropriate.  Thus, it becomes critical to have mechanisms through which requesters can discriminate between the interactions and labor that workers are expected to control (according to the policies of the market) and those that were outside workers' control \cite{dellarocas2005reputation}. This first type of interaction is known as “mission-critical” and the latter as “non-critical” \cite{xiong2003reputation}. Gig marketplaces have historically had difficulties in ensuring that requesters focus on evaluating mission-critical factors \cite{inproceedings,gaikwad2015daemo}. 
To help, practitioners and researchers have started to investigate different interfaces for facilitating the generation of more mission-critical reviews \cite{filippas2017reputation} which use drop-down lists to guide people on what metrics to focus on \cite{irani2013turkopticon}. However, these interfaces rarely focus on guiding people on the written reviews. As a consequence, unjust written reviews continue to plague gig markets \cite{Lu:2010:ESC:1772690.1772761, benson2019can} and have resulted in problematic outcomes, such as termination of workers' accounts. This can eliminate an important source of income \cite{rosenblat2016algorithmic}.

Seeing the need to motivate requesters to write fairer written reviews on gig markets, we present a new novel tool, Reputation Agent, which is an intelligent web plugin that detects when requesters have written reviews that consider factors outside a worker’s control. In such cases, Reputation Agent prompts people to reflect and focus on the performance metrics that are actually within the workers' control sphere. We designed Reputation Agent as a web plugin to empower platform maintainers to easily integrate the tool into their existing gig markets without having to change any of their front-end interfaces.  Fig \ref{fig:img01}. presents an overview of how Reputation Agent functions.  We conducted a study to evaluate how effective Reputation Agent was in prompting people to generate reviews that focus on metrics within workers' control. In order to investigate our tool in depth, we chose various gig markets (Uber, Upwork, and Grubhub) and recruited 480 reviewers to evaluate gig workers across several scenarios.
Across these different gig markets, reviewers working with Reputation Agent were motivated to focus significantly more on metrics that the worker could control instead of metrics outside of the worker’s scope.

Our paper contributes a new tool which leverages machine learning for fostering fairer written reviews about gig workers. Our design also provides a novel understanding of requesters who didn't change their reviews for the following reasons: truthfulness, empathy, warning, and agency. Through the ease of implementing our plug-in tool and the understanding gained about why requesters don't change their reviews, we hope that further studies can elaborate on the importance of integrating requesters' decision-making process into their studies to order to achieve fairer reviews for workers. Our discussion: {\em (i)} focuses on how tools such as Reputation Agent can motivate requesters to write more accurate performance reviews in a manner that provides productive feedback for the crowd market community; and {\em (ii)} explores how tools, like Reputation Agent, could help to develop empathy and motivate reflections on the type of policies and agencies that people desire within a gig market.
Our hope is that systems, like Reputation Agent, can initiate a future environment where workers operate in a fairer, more truthful space; an atmosphere in which all participants have a clearer understanding of the policies and labor conditions of the crowd market. This along with the hope that it will guide a future where gig workers no longer fear unjust termination.   

\section{Related Work}
The design of Reputation Agent is based on two main areas: (1) tools for written reviews; and (2) reputation systems.

\subsection{Gig Marketplaces}

Gig markets bring new jobs to the marketplace \cite{broughton2018experiences}. However, due to the nontraditional nature of the gig economy, criteria and tools to improve the labor conditions for workers are still necessary for researchers to investigate to ensure a safe and fair working environment for gig workers. \cite{berg2018digital,silberman2017rating, hara2018data, hitlin2016research,horton2011condition,irani2013turkopticon,irani2016stories,silberman2009fifteen,benson2019can,kaplan2018striving,harmon2018rating}. Gig markets rely heavily on reviews to help requesters identify which workers to hire and help workers ensure fair compensation \cite{kittur2013future,scholz2017uberworked}. It is this reliance on reviews that our study focuses on as bad reviews can pose an obstacle to workers. This is due to the fact that gig markets have been plagued with unfair reviews which contain inaccurate reputation signals about workers. These unfair reviews can ultimately limit workers’ future job opportunities and can also result in workers not getting paid or even being terminated from the marketplace. Unfair reviews are generally created because employers have a hard time differentiating the factors within the workers' control and the ones that have little to do with their performance (e.g., when they complain about an Uber driver getting stuck in traffic). However, because market power is typically placed in the hands of employers, a bad worker review can result in the worker losing her entire livelihood. It is important to research how tools can be implemented to protect gig workers \cite{williams2019perpetual}.

\subsection{Tools for Written Reviews}
Platforms for improving people's written reviews can be divided into two main types: Interface or Artificial Intelligence based.

\subsubsection{Interface Based.} Several interfaces have emerged that focus on driving people to provide better-written reviews about others. One set of these systems has focused on guiding better reviews within educational systems\cite{kulkarni2015peerstudio,cambre2018juxtapeer}. 
Cook et al. \cite{cook2019guiding} explored how the use of interfaces that have guiding questions can facilitate the generation of better reviews within project-based learning.
In our research, we build on the ideas behind these systems to now imagine interfaces that guide requesters to write fairer reviews about their fellow workers. 

\subsubsection{Artificial Intelligence Based.} Another subset of related tools have focused on using artificial intelligence to help reviewers. The work of Krause et al. \cite{krause2017critique} explored how natural language models could be used to guide designers to provide higher quality reviews about the work of their peers (which was not necessarily fair). Inspired by these ideas, we explored with Reputation Agent how different language models could be used to now guide requesters to write fairer reviews.  
For Reputation Agent we also used language models to identify when a requester is writing a review that is unfair and guide requesters to write reviews more focused on factors that workers controlled. 

Some of the first intelligent tools around reviews were automated methods that inferred the expected reputation scores that people would input based on their written reviews \cite{Alexandridis:2019:FUR:3308560.3316601, qu2010bag}. Others developed sentiment analysis methods to detect the polarity (positive, negative or neutral) of reviews \cite{Du:2016:ASW:2872518.2889403, collomb2014study}. Sentiment analysis has played an important role in improving the automated analysis and understanding of text reviews \cite{Luiz:2018:FSR:3178876.3186168, guzman2014users, elshenawy2016s, bartoli2016best}. Similarly, developments in deep learning algorithms have further facilitated the automated understanding and even categorization of marketplace reviews \cite{Kokkodis:2012:LPU:2187980.2188119, kumar2017fairjudge}. Deep learning algorithms and other related methods have facilitated automatically detecting more complex metrics aside from sentiment, such as the expected level of helpfulness of a review \cite{tang2013context}, who was to blame for a car accident based on a car insurgence report \cite{estival1995nlp}, health risks in restaurants based on people's reviews on Twitter \cite{sadilek2013nemesis} or detecting biased Amazon reviews \cite{elmurngi2018unfair}. We use inspiration from these intelligent systems to envision how deep learning could be used to automatically flag unfair reviews.

\subsubsection{Fairness In Crowd-Powered Text Reviews}

Within the context of Gig markets, fairness usually relates to the conditions of the workers laboring on these platforms \cite{fieseler2019unfairness}. Graham et al. \cite{graham2019fairwork} recently created a framework to score gig markets based on how fair they are to workers. Some of the variables considered were whether the platform paid gig workers the minimum wage and ensured their health and safety at work. Other measures revolved around whether the contracts and policies were transparent, concise, and accessible to workers.
Our goal with Reputation Agent, inspired by the latter point, was to facilitate mechanisms through which the policies of a gig market could be presented in a clearer, more conscientious manner. However, our focus was not just on presenting the policies to workers. But rather, facilitating an understanding of policies by requesters, who must judge workers and can, ultimately, have a lasting effect on their future job opportunities. Additionally, we focused on designing a tool that could be easily adapted to any gig market. We believe fairer marketplaces can be constructed by presenting more clearly to employers the roles of workers. 

Ensuring fairness in performance evaluations is a common challenge across gig markets \cite{borromeo2017fairness, mehrotra2018towards}. Unfair evaluations can come from individuals or groups \cite{allahbakhsh2014representation}. Several systems have implemented different mechanisms to ensure that the evaluations that people generate about others are fair. However,  most of these systems operate only at the score or metrics level \cite{schiffner2011privacy}.
These systems, generally, do not take any action to correct ``nasty''reviews. But, leaving unfair textual reviews intact can mean that the review can continue to affect the person long after the interaction took place. With Reputation Agent we focus on addressing this problem by detecting unfair reviews, and guiding employers to take action to correct them. 

\subsection{Reputation Systems}

A reputation system is any platform that evaluates businesses or peers based on an algorithm or customer rankings, ratings, or written comments \cite{zacharia2000trust}. The premise is to have parties rate each other which results in a score. This score should assist other parties in deciding whether or not to continue interacting with that party in the future \cite{josang2007survey}.
To operate effectively, reputation systems require at least three properties: long-lived entities that inspire an expectation of future interactions; capture and distribution of feedback about current interactions (information must be visible in the future); and use of feedback to guide trust decisions \cite{resnick2000reputation}. 

The end goal of reputation systems is to strengthen the quality of markets and communities by providing an incentive for good behavior and quality services, and by sanctioning bad behavior and low-quality services \cite{josang2009challenges}. In order to achieve that goal, some reputation systems have implemented diverse workflows and validations. PowerTrust \cite{zhou2007powertrust} takes the power-law distribution in user feedback to get a more accurate global reputation. Whitby et al. \cite{whitby2004filtering} describe a statistical filtering technique for excluding unfair ratings via a Bayesian reputation system.
Notice that prior work focused on improving score based reputations, while our work is based on the foundations of these systems to now assist gig markets in reducing the number of unfair written reputation signals.

\subsubsection{Reputation Systems For Gig Markets}

Within gig markets, reputation systems typically focus on evaluating the different actors involved in the market (workers, requesters and the platform itself)\cite{allahbakhsh2012reputation}. Reputation systems within the context of gig markets have become a key component for selecting the workers and clients with whom one will collaborate. A worker’s income positively correlates with higher reputation scores \cite{gandini2016reputation, horton2015reputation}. Therefore, ``bad reviews'' can affect worker's access to employment and can overall affect workers’ livelihood. Thus, designing accessible tools can promote fairness and enables a shift in the power dynamics \cite{williams2019perpetual}.

Different interfaces have been introduced to prompt and guide people to write reviews that better match the labor of workers and are potentially more fair and maintain more accurate reputations on the marketplace. For instance, Gaikwad et al. \cite{gaikwad2015daemo} developed Boomerang in Daemo Crowd Market, and explored interfaces that benefited requesters by sharing more accurate information about workers and penalizing requesters who shared inaccuracies. Such mechanisms might not only help workers to obtain better assessments of their work, but it can also help to address the ballot stuffing problem (where people get too many positive reviews, and it thus becomes difficult to assess who is ``good'').  Our research is inspired by these prior mechanisms to drive fairer reviews and prevent assessments that may unfairly affect the reputation and even the income of gig workers.

\section{Reputation Agent}

We argue that a way to enable fairer reviews in gig markets is via systems that can present requesters with transparent policies that pertain to workers without interrupting their review writing process. This information should only be highlighted in cases when the system identifies that the reviewer has included unfair factors (i.e., factors, per the market's policies, outside a worker’s control). For this purpose, our research explores: (1) machine learning techniques that detect when an individual is focusing her review on factors outside the workers’ sphere of control; and (2) interfaces that use that information to then prompt the person to reconsider her review in order to refocus on factors within the worker's control. Reputation Agent has two main components: a ‘Smart Validator’, to detect elements of a review that includes factors outside a worker’s control; and a ‘Fairness Promoter’, to guide people to focus their review on the factors that were within the worker’s control. Figure \ref{fig:img01} shows how Reputation Agent enhances existing review forms with these two main parts.

{\bf Smart Validator.} This component learns to detect when a review has factors outside the worker’s scope according to the policies of the market. The Smart Validator has an end-to-end workflow for training a machine learning model. The steps are:

A. PREPARE TRAINING DATA. This piece focuses on collecting real reviews about gig workers. It functions as a web crawler that collects data from websites, such as SiteJabber and ConsumerAffairs, that share real-world reviews about gig workers. Once the data is collected, the module focuses on labeling each review based on whether it focuses on mission-critical metrics or not (i.e., factors that the worker controlled or not)  The labeling is done by analyzing the policies of each gig market.

B. TRAIN AND TEST INTELLIGENT MODEL. Given a set of labeled reviews (i.e., reviews that are labeled as to whether they are fair or unfair), Reputation Agent uses stratified sampling to split the labeled data into training, test, and validation sets under proportions of 80\%, 10\%, and 10\% (the validation set helps to avoid overfitting).  Using Python 3 and the Keras framework with Tensorflow, we trained eight models to learn to recognize reviews that evaluate workers based on mission-critical metrics and non-critical ones. Our goal was to identify the machine learning models which worked the best for different gig markets. For this purpose, we trained different machine learning models which used as feature vectors either word vectors or embeddings:

{\bf Word ngram + LR}: Logistic regression with word ngrams.

{\bf Char ngram + LR}: Logistic regression with character ngrams.

{\bf (Word + Char ngram) + LR}: Logistic regression with word and character ngrams.

{\bf RNN no embedding}: Recurrent neural network (bidirectional GRU) without pre-trained embeddings.

{\bf RNN + GloVe embedding}: Recurrent neural network (bidirectional GRU) with GloVe pre-trained embeddings.

{\bf CNN (multi-channel)}: Multi-channel Convolutional Neural Network.

{\bf RNN + CNN}: Recurrent neural network (Bidirectional GRU) + Convolutional Neural Network.

{\bf Google BERT} \cite{devlin2018bert}: Bidirectional Encoder Representations from Transformers, is a new method of pre-training language representations which obtains state-of-the-art results on a wide range of Natural Language Processing (NLP) tasks.

\begin{figure}
\centering
  \includegraphics[width=0.9\columnwidth]{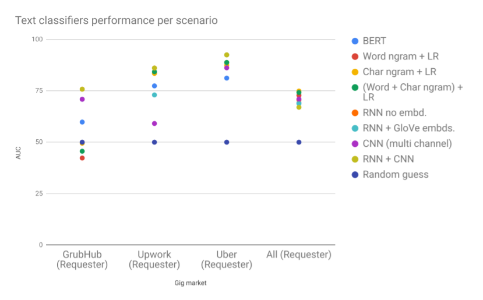}
  \caption{Text classifier benchmark. Recurrent Neural Networks (RNN) + Convolutional Neural Network (CNN) approach performed better across conditions.}
  \label{fig:img03}
\end{figure}

We implemented early stopping as a method to stop training once the model performance stops improving on a hold out validation dataset. For the deep learning models, we used a binary cross-entropy loss function,  ADAM as an optimizer, and a learning rate of 0.001. Fig. \ref{fig:img03} presents an overview of the benchmark of the training models (i.e., the figure shows the performance metrics of each model). We note that different machine learning models performed better on different gig markets. However, RNN (Recursive Neural Network, a Deep Learning Algorithm) performed in general the best across all gig markets. This was the reason we eventually choose to utilize this model. After the model has been trained, it is exposed as a REST web service via JSON requests to our front-end interface. The service is consumed directly by Reputation Agent’s Accuracy Promoter and it displays the messages accordingly. 

{\bf Fairness Promoter.} This component displays messages to the reviewer to prompt them to avoid considering factors outside the worker’s control based on the policies of the market. It displays the prompt messages that the platform maintainer defines and the messages are triggered based on the predictions from the Smart Validator. The Fairness Promoter is a web plugin for JQuery, a javascript framework, and works as a form validation plugin (commonly used to prevent forms from submitting data that do not fulfill a website’s validation or formatting criteria). Reputation Agent’s Fairness Promoter is linked to a text control that triggers a request to the Smart Validator every time the text control stops being used by the reviewer. 
The Fairness Promoter sends the reviewer’s current text to the Smart Validator in a JSON format. The Smart Validator analyzes the text and returns the likelihood of whether or not the review is focusing on factors that were within the control of the gig worker. If it is not, the Fairness Promoter then displays its configured messages to prompt reviewers to reconsider their review and focus instead on variables that the worker was able to control. 

In our design of the Fairness Promoter we chose for in-form prompting instead of popups. The logic behind this decision is that this design can lead to faster completion times \cite{hofseth2019form}. This is an important decision due to the limited time that customers usually spend evaluating services. Additionally, we considered that some users might have popup blockers in their browsers that could prevent them from seeing the prompt. Therefore, we opted to explore other approaches.  Additionally, we chose the prompted message to be shown after the end-user finished writing her review, instead of while she was completing it. We made this decision because prior research has shown that people tend to be in either a form-completion-mode or a problem resolution mode \cite{bargas2007usable}.  If people are in a form-completion-mode, they tend to ignore alert messages (and hence Reputation Agent would be less effective). Furthermore, we decided to place Reputation Agent’s prompting messages close to the review text box since previous work \cite{seckler2012user} has shown that such placement is more effective than when it is placed on top or at the bottom of the review form. On the other hand, while our prompting messages can be edited by platform maintainers to publish whatever message they desire; we aimed for the initial boxed messages to follow guidelines that prior work has deemed are the most effective. In particular, we follow the design guidelines from Bargas-Avila et al. \cite{bargas2010simple} that stated that promoting messages should be polite, explain the problem, and outline a solution.  Our explanation aimed to convey to requesters how their review might be considered unfair based on the policies of the market. We also aimed to briefly explain what type of factors are considered to be outside the control of a worker; and offered people the solution to re-write the parts of their review with those unfair factors.

\section{Evaluation}
Reputation Agent instantiates our design hypothesis that by flagging reviews with factors outside a worker’s control and then presenting to employers what the policies of the gig market highlight as workers’ responsibilities, we can prompt fairer assessments of gig workers. To test this hypothesis, we conducted a between-subjects study comparing Reputation Agent with control interfaces. We had participants evaluate a gig worker, given a scenario where the customer had experienced a “bad outcome” on the gig market. However, it was not the fault of the worker being evaluated (i.e., factors outside the worker’s control were to blame). We study whether people using Reputation Agent generated fairer reviews than people using control interfaces to review gig workers under the same circumstances. Given that it was also important for us to evaluate our tool within different gig markets (considering that it could be used in diverse niches), we evaluated our tool on marketplaces similar to: Uber, GrubHub and Upwork.  

\subsection{Method}
Our study focused on three popular gig markets (Grubhub, Uber, and Upwork). We randomized participants into one of our experimental conditions which represented a particular gig market and interface for reviewing workers (Reputation agent or control). Participants had to imagine they were a gig market customer or employer who had to write a review about a gig worker after experiencing a ``bad'' outcome on the marketplace (which was not the worker's fault). The scenarios that participants had to consider were:

{\bf 1) Uber Scenario.} Participants are passengers in a ride-sharing platform (e.g., Uber) where the driver had followed the recommended GPS route, had a clean car, had picked them up and dropped them off in the correct locations, and was polite. However, due to the heavy traffic, they experienced a delayed trip and had to pay an overpriced fee. The tardiness of their ride resulted in them missing an important meeting with a client and losing a contract.

{\bf 2) GrubHub Scenario.} Participants have an important lunch with a client and ordered the meal through an on-demand food delivery platform (GrubHub). The delivery person dropped off her meal on time. However, the meal contained an ingredient that caused the client to have an allergic reaction; thus, making the client very sick (The order had included a request for this ingredient to be removed). Due to the bad experience, the client decided to cancel her contract with our participants.

{\bf 3) Upwork Scenario.} Participants used a freelancing platform (Upwork) to hire someone to translate an important report from English to French for a French client. The translation was delivered on time and the translation seemed to be of high quality. However, due to a glitch in the system, the last part of the essay was truncated. Because the customer did not know French they had not realized that the report was truncated. They gave the truncated translation to their French client making a bad impression and losing the contract with the client.

For each of these three gig markets, we trained our tool to detect reviews that involved factors outside a worker’s control. For this purpose, for each gig market we: (1) collected 1,000 real-world reviews from SiteJabber (for the three scenarios); (2) had two independent college graduate coders classify each of these reviews into whether they involved worker’s performance or factors outside the worker’s control. We provided summaries of what factors were considered to be within the worker's control and examples of which ones were not.  Coders were also given a link to the policies of each of the three gig markets to better assess the variables that the marketplace considers are under a worker’s control. Some explained examples were given to coders to have a common agreement when dealing with ambiguous cases. The two coders agreed on the classification of 94.7\% of all the reviews (Cohen’s kappa =.86: Strong agreement). We then asked a third college graduate coder to act as a tiebreaker in cases of disagreement. After this step, for all three types of gig markets, we had a labeled set of reviews. The labeled data was provided as input to Reputation Agent’s Smart Validator to train its models. Reputation Agent uses stratified sampling to split the labeled data into training, test, and validation sets under proportions of 80\%, 10\%, and 10\% (the validation set helps to avoid overfitting). We implemented early stopping as a method to stop training once the model performance stops improving on a hold out validation dataset. For our deep learning module, we used a binary cross-entropy loss function,  ADAM \cite{kingma2014adam} as an optimizer, and a learning rate of 0.001. 

Across conditions, participants wrote their review based on their fictional scenario and using one of the four possible interfaces:

{\bf 1) Control (written text).} The end-user is presented with a traditional textbox where they must write a review about the worker.

{\bf 2) Control + Rating.} The end-user is presented with a traditional textbox where they must write a review about the worker, as well as complete traditional 5-star rating questions. These rating questions match the ratings that are currently present in the particular gig market in which the participant is operating (e.g., participants in the Uber scenario were presented with the rating questions that Uber uses to review drivers).

{\bf 3) Reputation Agent.} The end-user is presented with a traditional textbox where they must write a review about the gig worker while receiving prompting from Reputation Agent. 

{\bf 4) Reputation Agent + Rating.} The end-user is presented with a traditional textbox where they must write a review about the gig worker while receiving prompting from Reputation Agent. The end-user is also asked to complete traditional 5-star rating questions about the worker. Here, the rating questions again match the questions present in the given gig market. 

\begin{figure}
\centering
  \includegraphics[width=1.0\columnwidth]{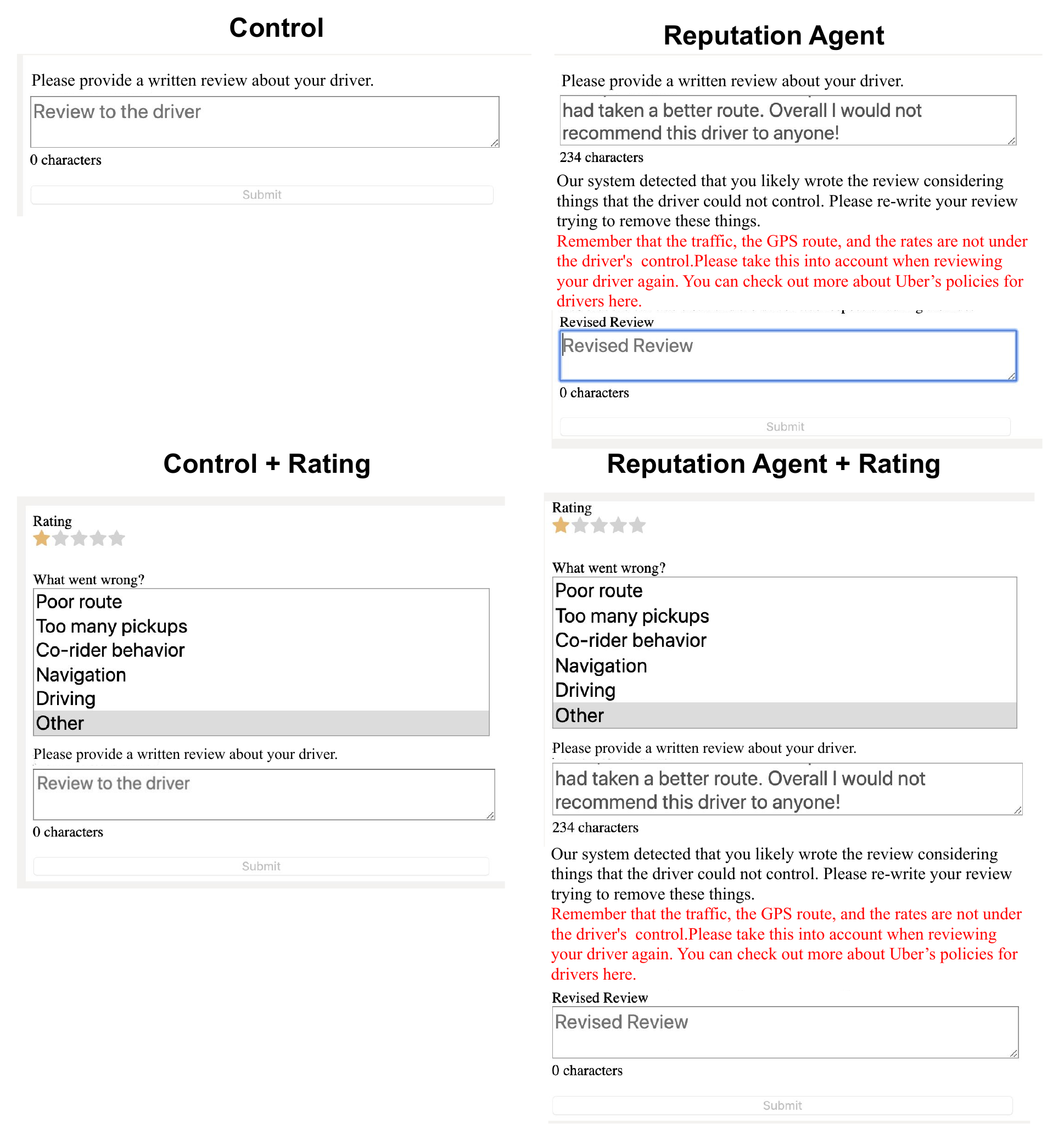}
  \caption{General Interfaces per condition.}
  \label{fig:img05}
\end{figure}

\begin{figure}
\centering
  \includegraphics[width=1.0\columnwidth]{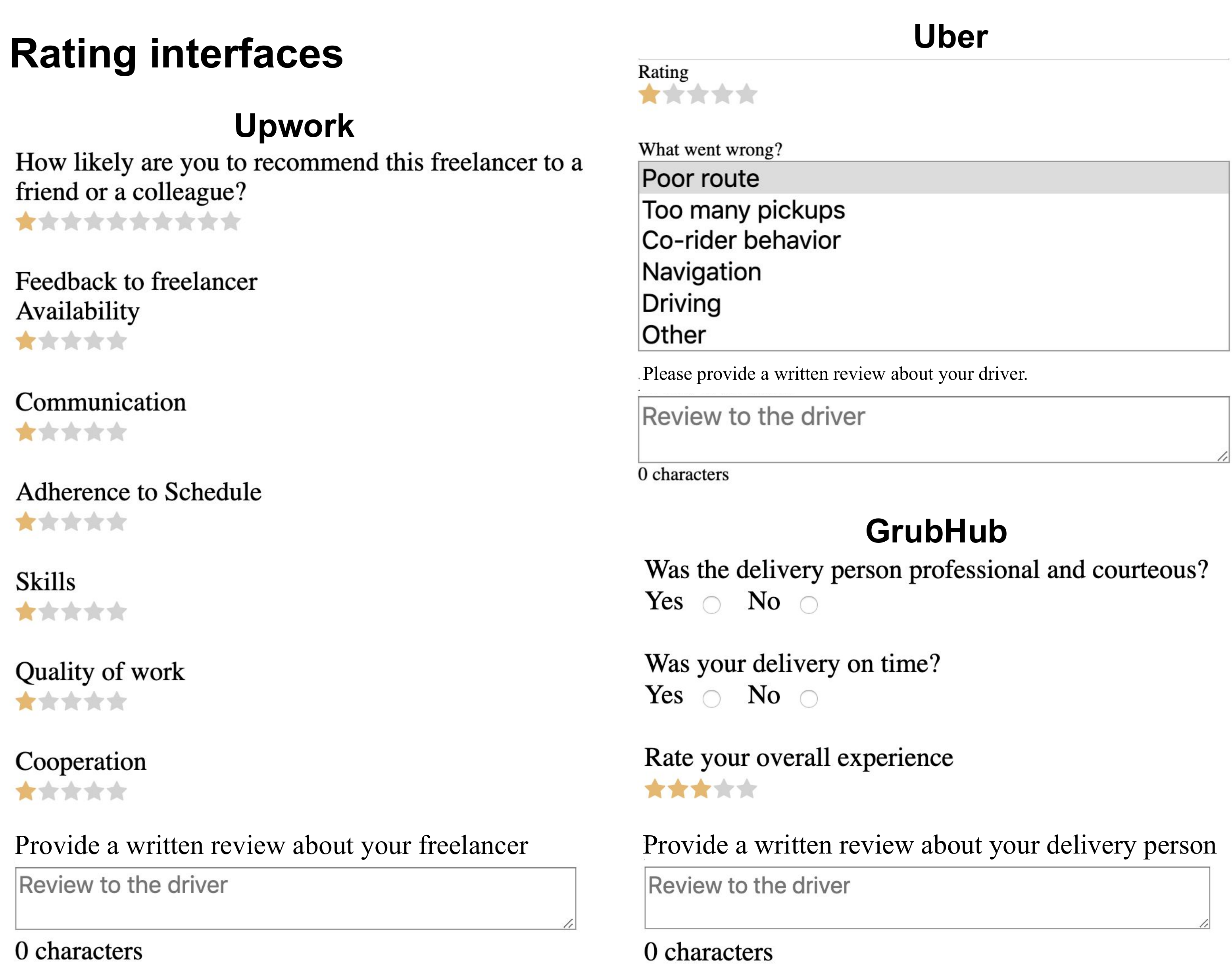}
  \caption{Rating interfaces per Gig Market.}
  \label{fig:img06}
\end{figure}

Fig. \ref{fig:img05} presents a general overview of how the interfaces per each condition looked. Fig. \ref{fig:img06} presents the different rating interfaces we considered per gig market. We aimed for these rating interfaces to mimic the ratings that particular gig markets have as we were interested in studying how our tool performed within mainstream settings. Notice that in the Reputation Agent conditions, we stored all the review attempts to analyze how people’s behavior changed.

Our between-subjects study had 12 conditions that involved three different fictional scenarios (three types of gig markets) where four different interfaces for reviewing workers were used. Each condition had 40 participants. People’s participation consisted of writing the review for the worker they were assigned and then completing a follow-up survey to provide feedback about their experiences. Specifically, the survey questioned people about: (1) Who or what was responsible for the bad service they had received on the gig platform? (gig worker, requester, platform algorithms, client, or other) (2) How much fault did each of those actors have? (3) How much did they think that their review would affect the worker’s reputation? (4) As a customer of gig platforms, what type of review interface (written or 5-star rating reviews) did they prefer? (5) As a worker or requester of gig markets, what type of reputation mechanism (written or 5-star rating reviews) did they prefer? (6) How much did they feel that the interface helped them to give more accurate feedback about worker’s performance? (7) How did their review process (if any) change after completing the review with their interface? Once participants had submitted their review and completed the follow-up survey, we analyzed whether the reviews they submitted were fair, specifically whether they integrated factors outside a gig worker’s control or not (to study the effectiveness of Reputation Agent). For this purpose, we had two independent college graduate coders read each of the final reviews that participants generated and categorize whether the review blamed the worker on factors outside the workers’ control or not (coders were also given the policies of each gig market to help their categorization, examples and summaries of the policies). The two coders agreed on the classification of 95.1\% of all the reviews produced by participants (Cohen’s kappa =.87: Strong agreement). In cases where there was disagreement, we asked a third college graduate coder to act as a tiebreaker. In all cases, we categorized the first and last reviews submitted in order to determine how much Reputation Agent lead people to change their reviewing behavior.

We recruited a total of 480 participants using university mailing lists, social media, and via postings on gig markets. Note that these are the same methods utilized by prior work to recruit requesters for studies \cite{gaikwad2015daemo}. Important to note is that all our participants had been at least once a requester (employer or customer) on the gig market to which they were assigned. 55\% of our participants were male, and 45\% were female. Participants were between the ages of 21 to 70 years old, with the median age being 35. All had at least a High School degree, 59\% had a bachelor's degree, 17\% a master's degree, and 2\% a Ph.D. degree. Some of our participants had been workers at least once on gig platforms: 18\% on Uber, 18\% on Upwork, and 12\% on GrubHub (which is normal given that gig markets allow people to take on both roles.) 
Participants were paid \$2.00 USD to participate, and the study took at most 15 minutes.

\section{Results}

\begin{figure}
\centering
  \includegraphics[width=1.0\columnwidth]{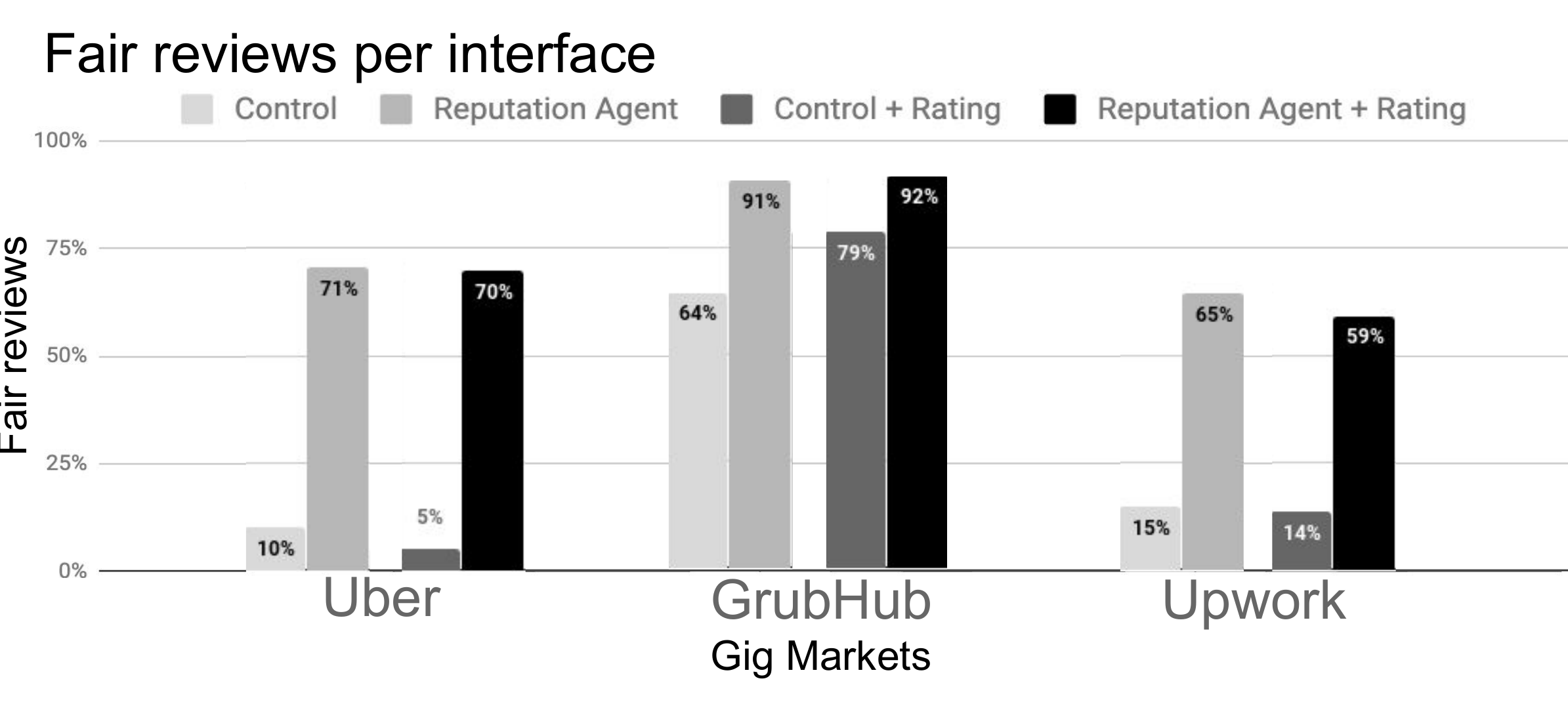}
  \caption{Percentage of fair reviews created per gig market scenario and when using a particular interface.}
  \label{fig:img07}
\end{figure}

Figure \ref{fig:img07} presents across gig markets the number of fair reviews that were generated when using a particular interface. Table \ref{table01} presents examples of reviews that were classified as unfair and fair. Across gig markets, the people using Reputation Agent wrote a larger number of fair reviews. In certain scenarios, having people use Reputation Agent leads to an increment of fairer reviews in comparison to when Reputation Agent was not used. For instance, for the Uber scenario when people used the control interface, only 10\% of the reviews were fair (i.e., 90\% were unfair reviews where people blamed their driver for factors outside her control, e.g., bad traffic.) However, when using Reputation Agent the number of fair reviews increased up to 70\%.

We also note that in some scenarios, having Reputation Agent operate with both textbox and numerical rankings lead people to write a slightly higher number of fairer reviews than when using only a textbox (this is the case for the GrubHub scenario). However, we also note that for Upwork and Uber there were a higher number of fair reviews when Reputation Agent operated only with a textbox (and no numerical ranking). To test whether these observed differences were significant or not, we conducted a series of statistical tests. After determining that our data did not meet the normality assumption, we decided to run an omnibus non-parametric Kruskal-Wallis test (p < .00001, H = 81.9303) and the Mann–Whitney U-tests with Bonferroni correction (p < .00001, z = 9.05126, U = 20400) to identify post hoc effects over conditions. Through this analysis, we found that there was indeed a significant difference in the number of fair reviews that people generated when using Reputation Agent when compared to reviews generated with traditional interfaces. In other words, we found that participants are significantly less likely to write unfair reviews when using Reputation Agent than with normal interfaces.

We also investigated how much Reputation Agent helped people to start changing their reviewing behavior. For this purpose, we measured the number of people who, while using Reputation Agent, changed their original review from being unfair to fair (see Table \ref{table02}). On average, across gig markets, Reputation Agent was able to convert two out of every three of the reviews that were originally unfair to fair (67.1\%). For when people decided to not change their unfair review, we analyzed the reasons for this behavior by observing what they stated in the survey. This analysis can help us to identify some of the challenges that Reputation Agent has in ensuring fairness in gig markets. We used open coding to extract initial concepts from people's responses \cite{mihas2019qualitative}. We aimed for these initial concepts to consider the themes that related work had derived on people's motivations for writing certain types of reviews \cite{goodman2017crowdsourcing, levitt2007laboratory}. Next, we discussed these initial concepts as a group to iterate on them and created a codebook (list of themes).

\begin{table}[]
\caption{Examples of reviews that were written by study participants and categorized as “unfair” and “fair” across gig market scenarios.}
\begin{tabular}{|p{0.6cm}|p{7.3cm}|}
\hline
{\bf \tiny Unfair}     & {\bf \tiny Blame worker for things outside their control}  \\ \hline
\tiny \tiny Uber & \tiny “I am beyond pissed. This driver took a ridiculous route causing a 45min delay and had the nerve to overcharge me on it! I was late for work and obviously, my boss was not pleased. Thanks a lot, Uber.” \\ \hline
\tiny GrubHub & \tiny “The delivery person failed to perform one of the basic functions of their job, which was ensuring the product they picked up was correct. He brought lunch to one of my clients and she couldn’t even eat it due to allergy concerns. Our food had peanut butter when I clearly stated no peanut butter. Very annoyed. I wasted my time and my money.” \\ \hline
\tiny Upwork & \tiny “Working with this person was a pain in the ass! I had a very important proposal to give to an important potential client who only spoke French. As I only speak English I contracted with this worker to translate it for me. They submitted too close to the deadline with little wiggle room, making it impossible to fix or check for any issues. They sent accidentally incomplete work. It cost me a contract due to his negligence. I do not recommend him, as he’s not meticulous.” \\ \hline
{\bf \tiny Fair}     & {\bf \tiny Assessed factors that the worker could control.}  \\ \hline
\tiny Uber & \tiny “The GPS of my driver leads me to a very congested route today and it took me a lot more time and money to get to work. But these things happen. I could have made the same choice driving myself so the driver can’t be blamed for something out of their control. Driver and his car were very nice. I am however very unhappy with the service and will be immediately unsubscribing soon.” \\ \hline
\tiny GrubHub & \tiny “The delivery person got the food to me on time. It was not their fault that the order was wrong. The order was wrong because of the people at the restaurant. They messed up what was suppose to be a great lunch. The driver did all he was supposed to do and I give him a good review for that..” \\ \hline
\tiny Upwork & \tiny “I thought the worker did a good job and the work was presented well, it is a shame however that the system failed at the last minute and I got an incomplete submission.” \\ \hline
\end{tabular}
\label{table01}
\end{table}

\begin{table}[]
\caption{Corrected reviews after Reputation Agent prompted participants.  2 out of every 3 reviews were corrected.}
\begin{tabular}{p{1.0cm}|p{4.0cm}|p{2.0cm}}
\tiny  Gig market               & \tiny Condition     & \tiny Corrected reviews \\ \hline
\tiny Uber    & \tiny Reputation Agent          &  \tiny 66.6\%   \\    
                         & \tiny Reputation Agent + Rating &  \tiny 70.2\%   \\  \hline		
\tiny GrubHub & \tiny Reputation Agent          &  \tiny 80.9\%  \\   	
                         & \tiny Reputation Agent + Rating &  \tiny 73.3\%   \\  \hline		
\tiny Upwork  & \tiny Reputation Agent          &  \tiny 57.8\%   \\   
                         & \tiny Reputation Agent + Rating &  \tiny 53.4\%   \\  
\end{tabular}
\label{table02}
\end{table}

The codebook with examples was shared with two coders that agree in the 91\% of the reasons (Cohen’s kappa =.81: Strong agreement) and a third college graduate coder to act as a tiebreaker in cases of disagreement. We detected the following categories describing participants' reasons for not changing their reviews:

\subsubsection{\bf Truthfulness.} Some reviewers (21\% of all reviewers who chose not to change their review after prompts from Reputation Agent) felt that changing their review implied lying about how they felt and wanted to keep their review as it was because it was truthful. Examples of their reasoning:

\emph{“I always give honest and detailed reviews and will continue to do so. This interface will not impact on my reviewing process. I will always stand by how I have done things in the past: with the truth!”} Uber Reviewer 44.

\emph{“Why on earth would my process change? [...] if I get bad work, I'm going to leave an honest review regardless of whether it was the "platform" or the worker's fault…”}, Grubhub Reviewer 15.

\emph{“...I like being honest and factual with reviews.  That’s why I don’t like changing my reviews…”} Upwork Review 43.

\subsubsection{\bf Agency.} These reviewers (25\%) felt that although a gig market’s policies might dictate that certain actors were not to blame, they believed that such actors should have had more agency in their decisions despite the policies of the gig market. 

\emph{“Drivers should be able to tell which routes are clean by instinct based on the day and the time of the day without even looking at the GPS. The drivers should know the city that they are driving very well...”} Uber Reviewer 23.

\emph{“Well the delivery guy didn't listen to a word I said so now my client can’t eat the meal. If he has any type of peanut in their food they can go into anaphylactic shock and die. That is not what I want for their lunch? Is that what you want them to have for lunch? Death? The delivery guy really needs to be responsible for this.”} Grubhub Reviewer 32.

\emph{“...Let’s be honest, the worker did not perform as well as expected. She missed that some words were cut off. The bottom line is that she needs to learn to handle herself responsibly in the world.”} Upwork Reviewer 19.

\subsubsection{\bf Empathy.} These reviewers (36\%) felt they needed more time to analyze the scenario before changing their review. They appeared to have empathy for all actors involved and wanted to truly understand their situation before changing their review.

\emph{“I like to think about all the circumstances before writing reviews. I like to use empathy. In my future review I will probably be a bit less harsh on drivers. I will think about the driver and how they treated me as well. But before I make those change I will try to calm down first…”}  Uber Reviewer 21.

\emph{“...I wanted to focus on the bad aspects of my meal, but then I realized I was supposed to focus on the deliverer only. So I switched it and focused on the delivery worker  instead.  I try to take all factors I am aware of into account when writing my review, and wait a bit so that any emotions associated with the work would not affect my review.  I think I would wait to re-write my review so I am not angry and really think about the worker…”} Grubhub Reviewer 5.

\emph{“I like to take software and platform issues into account before completely blasting a worker in a review (either with stars or a written review). I won’t change my review now because I have to stop and think about things. Instead of just getting mad and going off on the worker immediately. I would also want to have more communication between worker and purchaser before the review process, and have a way to discuss the review if either party truly found it to be in error.”} Upwork Reviewer 39.

\subsubsection{\bf Warning.} Some participants (13\%) maintained their review because it was important for them to have a space where they could caution others about what they experienced. They did not care about whether they blamed the incorrect person.

\emph{“...I just detailed the problems I encountered with the driver. I wanted others to know about my issue so that it doesn’t happen to them as well.”}  Uber Reviewer 49.

\emph{“...I usually never provide reviews for delivery people [...]  But in this case the service was exceptionally bad and my experience would serve as a cautionary tale to others. So that is why I can’t change it [the review]”} GrubHub Reviewer 5.

\emph{“...I think my review would let people know of the risks about using this worker/platform.  They could potentially avoid situations like the one that I was in. It’s important for me to keep my review to warn others....”} Upwork Reviewer 20.

67\% of the people using Reputation Agent reported that they felt that the interface helped them to be more aware of inaccuracies in their reviews. Participants across conditions reported that they felt their review would affect worker’s reputation. People in the control condition had the perception that their review would be the most harmful (mean 3.9 of 5). This is notable when compared to people using Reputation Agent who on average thought that their review would not be as harmful (mean 2.9 of 5).

\section{Discussion}
Our experiments demonstrated the potential of using intelligent web plugins to detect unfair reviews on gig markets, and then prompt fairer assessments by presenting micro-information about gig workers' conditions and policies. Across different marketplaces, the majority of people using Reputation Agent ended up writing fairer reviews. Our study provides a novel insight into how marketplaces could use this type of smart web plugin to bring more fairness to workers. In this section, we discuss opportunities and challenges we see with Reputation Agent, and highlight design implications for future systems that operate within the gig marketplace.

\subsection{Building Empathy In Gig Markets}

Taking empathy into account in the human-centered design process can align designers with the values and needs of people who may use the platforms \cite{10.1145/3290605.3300528}. Mencl et al. defines empathy as ``a positive moral emotion that aids reasoning\cite{mencl2009effects}.'' Our study highlighted that prompting people to reconsider their reviews and reason more deeply about the worker and what her actual job was, helped reviewers to be fairer.

While all our participants considered that their reviews would have an impact on workers' lives, the level of harm that people attributed to their review varied across conditions. People using the control (text only) interface tended to believe their reviews were the most harmful while people using the Reputation Agent + Rating interface felt they were doing the least harm to workers. The ``tension between reason and emotion when making decisions\cite{frith2008role}'' allows us to see the benefit of a tool such as Reputation Agent in prompting requesters to reconsider their written review. Thus, our results highlight that providing more metrics and guidance helped people feel as if they were doing less harm to workers while still submitting a review that was accurate.

We see tools like Reputation Agent as a way to help requesters have a more humane perspective of workers by providing more transparency and awareness of what the current labor conditions are in gig markets. Through Reputation Agent we offer a way in which requesters can be guided to better understand the actual job expectations for workers. 
We believe that through this transparency and highlighting of boundaries denoting gig workers' labor that we will be able to build more consideration for workers within gig markets \cite{irani2016stories}. 
Several of our participants who Reputation Agent prompted to change their reviews discussed how the tool helped them to better understand workers' conditions.

From our study, we also identified that there were cases where people even after being prompted by Reputation Agent, refused to change their review at all. Many of these individuals were people who felt that workers needed to have more agency. For instance, in the Uber case, some passengers believed that their Uber driver should not have followed the recommended GPS route, but instead selected a shortcut and better route. These individuals blamed their Uber drivers for not taking the initiative and knowing enough of their city to understand that the GPS algorithm was wrong. We believe that in these cases, it might be worth designing interventions where the policies and responsibilities of gig workers are explained in detail to these individuals from the outset. We believe there is an opportunity in using systems like Reputation Agent as a way to create more empathy between requesters and workers. Additionally, it might be worth explaining to workers the perspectives of these requesters in order to facilitate their understanding of why certain requesters might expect them to not always follow an algorithm and be more ``proactive.'' In these cases, we visualize platforms that do not penalize workers for not being proactive (i.e., by following the instructions of the algorithm), but rather help workers open their minds to other perspectives and help them to see that having more agency in their decision process could provide growth opportunities, e.g., to eventually become a manager. 

\subsection{Supporting Reflection In Gig Markets}

Our study contained 46 individuals who refused to modify their review even after being prompted. We considered it important to understand the reasons these individuals had for not changing their written assessments. In some cases, requesters used the review process as a chance to communicate to the platform that it would be more efficient (in terms of time and cost) if the worker was allowed more agency. We believe there are benefits to gig markets when they understand the type of agency that requesters want to see in the market.
Platforms could consider alternate methods/systems for capturing this type of feedback in order to protect workers.

Williams et al. found that tools that are based on only distributing ratings and reviews for task choosing decisions usually tend to create fragmentation and discrimination affecting the platform's fairness \cite{williams2019perpetual}. We argue that it is important to tie tools like Reputation Agent with platforms focused on driving citizen discussions and citizen reflection \cite{mahyar2018communitycrit}. 
On this point, it is important to consider the findings of Li et al. concerning the influence that embodied conversational agents (ECAs) have on persuading people to consider feedback that is offered them \cite{li2007my}. ``The use of agents that resemble users'' might be the necessary factor that allows requesters to consider the promptings of the Reputation Agent to be more valid. We must also ask: what type of agencies should we expect from the different actors of the market and why is it important that we expect such agency from them? Further research is needed to investigate the type of interfaces and workflows that could be used to incentivize and guide quality reflections about what people expect from workers, requesters, and the different policies of a marketplace. This type of system could uncover pain points that exist in current crowd markets and where policy changes might be needed. Our study also highlighted another reason why requesters did not desire to change their review: the importance for them to use the space to truthfully share their experiences. In the widely cited paper “The Market for Goods and the Market for Ideas” \cite{coase1974market}, it is argued that in the market for goods (i.e., the market where consumer goods and services are exchanged), government regulation is desirable; whereas in the market for ideas (i.e.,  the market where opinions or beliefs are interchanged), government regulation should be limited. Online reviews can be seen as something that delivers both  “goods” and “ideas”. On one hand, having a person write an objective review of the work someone did could be seen as if they are delivering a good. The good, in this case, corresponds to the overall assessment of the labor that the worker did. This assessment not only helps the market better contextualize and measure the labor that is being produced \cite{jagabathula2014reputation}, it can also boost the SEO of the marketplace \cite{shenoy2016ranking}. Thus, helping it appear higher in the results of search engines and ultimately bring in more customers \cite{rognerud2008ultimate}. The review can also help the worker get better credentials, access higher pay, and more requesters (i.e., the review might persuade other requesters to hire the worker). Reviews as goods deliver services to the marketplace, workers, and even other requesters. However, reviews also have the capacity to deliver opinions and beliefs, and hence can also belong within the market of ideas. Thus, we see value in being able to actively regulate the activities that belong to the market for goods, while permitting freedom of expression for activities that relate to the market for ideas \cite{coase1974market}. Reputation Agent offers an advancement towards this area by providing a way to regulate reviews within the market for goods and flagging reviews that might pertain more within the markets of ideas. Future work could pursue this avenue to design review interfaces to express both forms of reviews. 

\subsection{Rating Systems and Fairness}

The importance of rating systems and fairness is an essential element in gig markets, whether it pertains to rating workers or rating requesters\cite{silberman2017rating, irani2013turkopticon, dow2012shepherding}. Creating a fair working environment with structures designed to protect workers' rights to receive fair compensation for their labor ensures the reputation and success of gig markets\cite{felstiner2011working,adam2016digitalisation, metall2016frankfurt,benner2014amazonisierung,silberman2017rating} Thus, devising a tool for gig markets to implement in order to ensure fair reviews for workers brings us one step closer to achieving this.

To this purpose, our findings that people tended to write fairer reviews with Reputation Agent when working with the written interface is an important addition to the tools available to gig markets. In our study, we also discovered that people tended to write a larger number of unfair reviews when Reputation Agent was tied with numerical ratings. This was specifically the case with Uber and Upwork, where having numerical ratings tied with a written review, led to a larger number of unfair reviews than when working with just Reputation Agent and a written interface. Upon closer inspection, we identified that the problem was the fact that the rating systems of these gig markets did not distinguish worker’s performance from factors outside the worker’s control. For instance, when assessing a driver’s rating on Uber, the market provides a list of possible issues and presents “poor route” as an option even though Uber policies outline that drivers should always follow the recommend GPS route (unless explicitly instructed otherwise by the passenger). As a result, several participants selected “poor route” as an issue and then wrote lengthy reviews blaming the driver for the traffic (despite the prompts from Reputation Agent). Similarly, we noted that markets which differentiated between metrics pertaining to workers vs the platform, led people to generate fairer reviews. For instance, GrubHub has a rating system that differentiates between these two types of metrics, and we saw there was a decrease in the number of unfair reviews generated by participants. We see then the necessity of gig markets to not only incorporate tools that promote fairer reviews, but they themselves must also clarify and communicate the metrics that pertain to the workers.

Unfair reviews may also be the product of biases which do not necessarily reflect a worker's performance, i.e., when a worker gets more positive ratings than expected given the service she provided. These types of reviews can be influenced by cognitive biases such as confirmation bias \cite{klayman1995varieties}, driven by having prior beliefs; anchoring effect \cite{caputo2013literature}, relying more on the first piece of information offered and hence the current performance does not matter; or perception bias \cite{greenberg1991motivation}, motivated by how others might perceive you as the reviewer. Reputation Agent provides the opportunity to educate people about possible biases they might have and how those might be impacting their reviews. Here, we envisage that Reputation Agent’s prompts might provide information about biases in addition to information about the policies of gig markets. Future research could focus on personalized feedback according to personality or cultural biases that might exist \cite{krzystofiak1988implicit,martell1993effects,hogan1987effects,zwikael2005cultural,li2001we}. 

\subsection{Feasibility And Maintainability}

Through our controlled experiments, we identified that Reputation Agent was able to lead requesters to generate fairer reviews than when they worked with the control interfaces. However, to accomplish these results, it was necessary to have labeled data for each gig market on what constitutes fair reviews and what constitutes unfair reviews. 
While the labeled data sample that we used was relatively small in comparison to the large number of reviews that are generated on these marketplaces daily \cite{baj2017sentiment}, it is possible that new gig markets might have a difficult time collecting and labeling review data for Reputation Agent.  

We have released our system\footnote{https://research.hcilab.ml/reputationagent} to help gig markets easily adopt and use our tool. 
Additionally given that Reputation Agent can be easily implemented as a validation module, Platform maintainers could change their front-end review interface without having to worry about Reputation Agent suddenly not working. Reputation Agent’s deep learning nature makes it so that if a gig market changes its policies, platform maintainers with minimum knowledge in artificial intelligence can easily re-train Reputation Agent to be updated with the changes \cite{fandango2018mastering,toxtli2018expertwin}. In our website we have shared training examples for Reputation Agent's learning module so that website maintainers can easily start using our tool. 

\subsection{Key Design Considerations}

Our investigation unraveled design considerations for technology to support the generation of fairer interactions on gig markets. 

\subsubsection{\bf Tools for Learning about Gig Market Policies.} Reputation System can be seen as a tool that helps highlight the policies of a gig market. For instance, when people are writing a review for Uber, Reputation Agent shares Uber’s policies. Future work could explore how heuristics and hard-coded rules can lead people to better understand the policies of a marketplace and comprehend what falls under worker vs. platform responsibilities. The visualization of different privacy policy representations can improve the understanding of the different actors \cite{10.1145/1753326.1753492}.

\subsubsection{\bf Tools for Better Moderation.} Integrating artificial intelligence (AI) into gig markets can go beyond flagging unfair reviews \cite{dai2011artificial}. AI can also be intermixed with human moderators to facilitate a better understanding of the perspectives of requesters. For instance, given that Reputation Agent is able to store all the review attempts that people make, the system could detect cases when even after prompting the end-user to reconsider her review she still kept everything the same. In which case, the system could trigger an alert to human moderators to take a closer look at the review. We view Reputation Agent as tool that can alleviate moderators' labor. The sustainability and self-management of the tool also depend on the neutrality of the training data. Human-in-the-loop mechanisms can allow different actors to have agency giving everyone decision power, not just the few who can code \cite{williams2019perpetual} to define how the tool is learning and taking the decisions.

\section{LIMITATIONS and Future Work}
The insights from this work are limited by the methodology and population we studied. Our controlled experiment allowed us to begin understanding how users engage with Reputation Agent. Although, we cannot extrapolate on how people would respond if this approach were implemented in a field deployment with conditions such as limited time and reduced willingness to reconsider their reviews.
Our attempt to counter this issue was by implementing interfaces and creating scenarios that mimicked various gig markets and circumstances. However, future work could benefit from analyzing how systems like Reputation Agent are used when people are on the go and suffer from time constraints. 
While the scenarios we studied resembled very specific real-world situations, our results might not yet generalize to populations at large or to different types of situations. Further analysis is needed to understand how studies that leverage real gig market actors and Reputation Agent play out in helping users to give more objective reviews. 

Reputation Agent was designed to limit the amount of extra interface controls that platform maintainers would have to implement. The aim of this work is to provide a smart validation mechanism for existing interfaces, i.e., easy to implement and not invasive. Future work could also explore how adaptations in the workflow and interface controls, such as a separated textbox for the reviews that are generated with Reputation Agent, could lead to reducing unfair reviews. This work explored the effect of using Reputation Agent in two settings: with ratings tied to text reviews and with just written text reviews (without any ratings). We studied whether in these settings Reputation Agent could guide people to change their reviews to fair ones. We choose to focus on written reviews in which having one bad written review could not only lead to a worker having her reputation jeopardized, but also having her account terminated. Future work could explore how integrating fairness validators might also influence the numerical ratings that people give to workers. Our work replicates the current conditions of gig markets, where people are never initially prompted or reminded to be fair in their reviews, i.e., Reputation Agent prompts only when unfair reviews are given. We established this setting because we considered that customers would likely be busy individuals who simply wanted their service delivered. Therefore, constant reminders of the gig market's policies could be considered invasive.
If they have not written an unfair review, it might not need highlighting. Future work could explore how promoting fairness throughout different points in time (e.g., directly when starting to write the review or at the end) can lead to fairer reviews. Our study may also have novelty effects that need to be studied through longitudinal studies. Future work could explore how longitudinal studies can promote fair reviews over time. This was a controlled experiment and not a deployment, i.e., there was never money at stake and no real harm done to the worker. Future work can compare how our results differ from deployments in the real world.

{\bf Acknowledgements.} {Special thanks to Amy Ruckes, Ben Hanrahan, and Six Silberman for the immense feedback and iterations on this work. This work was partially supported by NSF grant FW-HTF-19541.}

\bibliographystyle{ACM-Reference-Format}
\bibliography{main}

\end{document}